\documentstyle[12pt,epsfig]{article}

\newcommand{\wh}{\widehat}
\newcommand{\br}{{\bf r}}
\def\beq{\begin{equation}}
\def\eeq{\end{equation}}

\begin{document}

\title{Theory of doorway states for one-nucleon transfer
reactions}
\author{B.L. Birbrair\footnote{E-mail: birbrair@thd.pnpi.spb.ru}
and V.I.  Ryazanov\\ Petersburg Nuclear Physics Institute\\
Gatchina, St.Petersburg 188350, Russia } \date{} \maketitle

\begin{center}
A b s t r a c t
\end{center}

The doorway states under consideration are eigenstates of the
hamiltonian which is the sum of the kinetic energy and the
infinite energy limit of the single-\-particle mass operator.
Only Hartree diagrams with the free-\-space nucleon--nucleon
forces contribute to this limit, and therefore the observed
doorway state energies carry an important information about both
the nuclear structure and the free-\-space nucleon-\-nucleon
interaction.

\section{Introduction}

The experimental data on the quasielastic knockout reactions
$(p,2p)$, ($p,pn$), $(e,e'p)$ etc. leading to the strongly bound
hole states of complex nuclei carry an important information
about both the nuclear structure and the free-\-space
nucleon-\-nucleon forces. Of course such information is
contained in all nuclear data, but these ones are distinguished
by the fact that the above information is obtained by simple
means thus being highly reliable. The reasons are as follows.

1. As shown by M.Baranger \cite{1} the doorway states for the
one-\-nucleon transfer reactions are eigenstates of nucleon in
the static nuclear field, see Sect.2, thus being solutions of
the problem of particle motion in the central potential well.
This is one of the most simple problems of quantum mechanics.

2. As a consequence of contemporary ideas about the $NN$
interaction mechanism, see Sect.3, the only contribution to the
static field of nucleus is provided by the Hartree diagrams with
the free-\-space nucleon-\-nucleon forces: the two-\-particle,
Fig.1a, three-\-particle, Fig.1b, four-particle, Fig.1c, etc.

\begin{figure}
\centerline{\vspace{0.2cm}\hspace{0.1cm}\epsfig{file=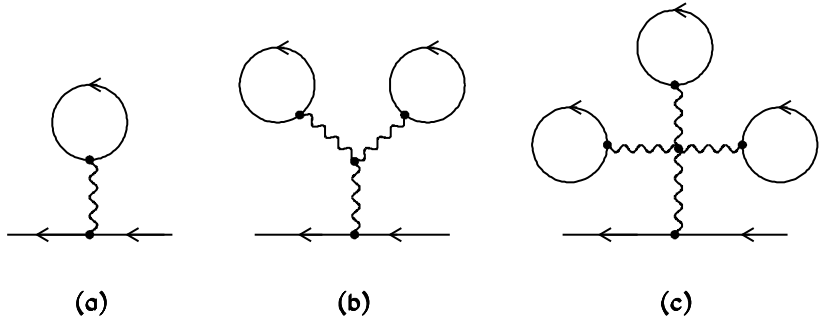,width=12cm}}
\vspace*{-0.5cm}
\caption{ Hartree diagrams for the static field of nucleus.}
\end{figure}

3. The two-particle contribution of Fig.1a is the convolution of
the free--space two--particle $NN$ interaction with the
nucleon density distribution in nucleus, and therefore it can be
determined from experiment. Indeed, the two--particle forces
are determined by the properties of deuteron and elastic $NN$
scattering phase shifts below the pion production threshold,
whereas the nucleon density distributions are deduced from the
combined analysis of the electron-\-nucleus \cite{2} and
proton-\-nucleus elastic scattering data \cite{3}.

The information about the many-particle contributions to the
static nuclear field (hence, about the free-\-space
many-\-particle $NN$ forces) can be obtained by comparing the
observed doorway state energies with the calculations including
the two-\-particle contribution only. In this way we found that
the free-\-space many-\-particle interaction includes at least
the three-\-particle repulsion and four-\-particle attraction,
see Sect.4.

\section{The Baranger theorem}

Evolution of the state arising from a sudden creation of
particle or hole in the ground state of nucleus $A$ is described
by the single-\-particle propagator \cite{4}
\begin{eqnarray}
&& S(x,x';\tau)\ =\ -i\langle A_0|T\psi(x,\tau)\psi^+(x',0)|A_0
\rangle\ = \nonumber\\
=&& i\theta(-\tau)\sum_j\Psi_j(x)\Psi^+_j(x')e^{-iE_j\tau}
-i\theta(\tau)\sum_k\Psi_k(x)\Psi^+_k(x')e^{-iE_k\tau}
\end{eqnarray}
with
\begin{eqnarray}
\Psi_j(x)=\langle (A-1)_j|\psi(x)|A_0\rangle\ , &&
E_j={\cal E}_0(A)-{\cal E}_j(A-1) \nonumber\\
\Psi_k(x)=\langle A_0|\psi(x)|(A+1)_k\rangle\ , &&
E_k= {\cal E}_k(A+1)-{\cal E}_0(A)\ .
\end{eqnarray}
So the propagator describes the evolution of the hole (particle)
state at negative (positive) $\tau$ values. According to Eq.(2)
the excitation energy region for the $A-1$ nucleus is
\beq
-\infty\ <\ E_j\ \le\ {\cal E}_0(A)-{\cal E}_g(A-1)
\eeq
(${\cal E}_g(A-1)$ and ${\cal E}_0(A)$ are the
ground-state energies of $A-1$ and $A$ nuclei), whereas that for
the $A+1$ nucleus is
\beq
{\cal E}_g(A+1)-{\cal E}_0(A)\ \le\ E_k\ <\ \infty\ .
\eeq
Such energy scale is convenient for us because the two regions
do not overlap in stable nuclei.

The Fourier transform of the propagator
\begin{eqnarray}
&& G(x,x';\varepsilon)\ = \int\limits^{+\infty}_{-\infty}
S(x,x';\tau)e^{i\varepsilon\tau}d\tau\ = \nonumber\\
=&& \sum_j\frac{\Psi_j(x)\Psi^+_j(x')}{\varepsilon-E_j-i\delta}
+\sum_k\frac{\Psi_k(x)\Psi^+_k(x')}{\varepsilon-E_k+i\delta}\ ,
\end{eqnarray}
(which is referred to as the single-particle Green function)
obeys the Dyson equation
\beq
\varepsilon G(x,x';\varepsilon)\ =\ \delta(x-x')+\wh k_xG(x,x';
\varepsilon)+\int M(x,x_1;\varepsilon)G(x_1,x';\varepsilon)dx_1\
, \eeq
$\wh k_x$ is the kinetic energy and $M(x,x';\varepsilon)$ is the
mass operator. The latter has the following general form
\beq
M(x,x';\varepsilon)\ =\ U(x,x')+\sum(x,x';\varepsilon)\ ,
\eeq
where the energy-independent part $U(x,x')$ is the static field
of nucleus, and the energy-dependent one $\sum(x,x';\varepsilon
)$ is responsible for all kinds of the correlation effects
(Pauli, particle-\-particle, particle-\-hole, ground-\-state,
long-\-rangle, short-\-range etc.). It has the following
high-\-energy asymptotics \cite{5,6}
\beq
\sum_{\varepsilon\to\infty}(x,x';\varepsilon)\ =\
\frac{\Pi(x,x')}\varepsilon\ +\ \cdots
\eeq
(the dots in the rhs denote the higher-power terms in respect of
$\varepsilon^{-1}$). As a result the static nuclear field is the
infinite energy limit of the mass operator,
\beq
U(x,x')\ =\ \lim\limits_{\varepsilon\to\infty}
M(x,x';\varepsilon)
\eeq
the decomposition (7) thus being unambiguous.

Now let us introduce the single-particle Hamiltonian
\beq
{\cal H}_{sp}(x,x')\ =\ \wh k_x\delta(x-x')+U(x,x')
\eeq
and its eigenstates
\beq
\varepsilon_\lambda\psi_\lambda(x)\ =\ \int{\cal H}_{sp}
(x,x')\psi_\lambda(x')dx'\ ,
\eeq
which are those of nucleon in the static field of nucleus. They
are not directly observed because they are described by only a
part of the total nuclear Hamiltonian. Their physical meaning
is however understood on the basis of the Heisenberg relation
according to which the infinite $\varepsilon$ value is
equivalent to the infinitely short time interval $\tau$. Hence,
the eigenstates of ${\cal H}_{sp}$, Eq.(10), describe the very
beginning of the evolution process under consideration thus
being the doorway states for the one-nucleon transfer reactions.

To demonstrate this more explicitly let us use the high-energy
asymptotics of the Green function (5):
\beq
G(x,x';\varepsilon)_{\varepsilon\to\infty}\ =\
\frac{I_0(x,x')}\varepsilon +\frac{I_1(x,x')}{\varepsilon^2}
+\frac{I_2(x,x')}{\varepsilon^3}\ +\ \cdots\ ,
\eeq
where
\begin{eqnarray}
&& I_0(x,x')\ =\sum_j\Psi_j(x)\Psi^+_j(x')+\sum_k\Psi_k(x)
\Psi^+_k(x')\\
&& I_1(x,x')\ =\ \sum_jE_j\Psi_j(x)\Psi^+_j(x')+\sum_kE_k
\Psi_k(x)\Psi^+_k(x')\\
&& I_2(x,x')\ =\sum_jE^2_j\Psi_j(x)\Psi^+_j(x')+\sum_k
E^2_k\Psi_k(x)\Psi^+_k(x')\ .
\end{eqnarray}
As follows from the spectral representation of the propagator,
Eq.(1),
$$
I_0(x,x')\ =\ i\bigg(S(x,x';+0)-S(x,x';-0)\bigg) \eqno{(13a)}
$$
$$
I_1(x,x')\ =\ -\bigg(\dot S(x,x';+0)-\dot S(x,x';-0)\bigg)
\eqno{(14a)}
$$
$$
I_2(x,x')\ =\ -i\bigg(\ddot S(x,x';+0)-\ddot S(x,x';-0)\bigg)
\eqno{(15a)}
$$
the above sums thus describing the beginning of the evolution
process $\left(\!\dot S\!=\frac{\partial
S}{\partial\tau}\right)\!,$ $\left(\ddot
S=\frac{\partial^2S}{\partial\tau^2}\right)$. Using the
definition (10) and the asymptotics (8) the Dyson equation (6)
may be written in the form \beq \varepsilon
G(x,x';\varepsilon)=\delta(x-x')+\int\left({\cal H}_{sp}(x,x_1)
+\frac{\Pi(x,x_1)}\varepsilon +\cdots\right)
G(x_1,x';\varepsilon)dx_1\ .
\eeq
Putting Eq.(12) into Eq.(16) and equating the coefficients at
the same powers of $\varepsilon^{-1}$ we get
$$
\sum_j\Psi_j(x)\Psi^+_j(x')+\sum_k\Psi_k(x)\Psi^+_k(x')\ =\
\delta(x-x') \eqno{(13b)}
$$
$$
\sum_jE_j\Psi_j(x)\Psi^+_j(x')+\sum_kE_k\Psi_k(x)\Psi^+_k(x')\
=\ {\cal H}_{sp}(x,x')  \eqno{(14b)}
$$
$$
\sum_jE^2_j\Psi_j(x)\Psi^+_j(x')+\sum_kE^2_k\Psi_k(x)
\Psi^+_k(x')\ =\ {\cal H}_{sp}^2(x,x')+\Pi(x,x')\ . \eqno{(15b)}
$$
As follows from Eqs. (14a) and (14b)
\beq
{\cal H}_{sp}(x,x')\ =\ -\bigg(\dot S(x,x';+0)-\dot S(x,x';-0)
\bigg).
\eeq
So the evolution of the hole (particle) state begins with the
formation of the nucleon eigenstates in the static field of
nucleus, the Baranger theorem thus being proved.

Now let us discuss the determination of the doorway state
energies $\varepsilon_\lambda$, Eq.(11), from the experimental
data. The weights of the doorway component in the actual nuclear
states are
\beq
s^{(\lambda)}_{j,k}\ =\ \left|\int\psi^+_\lambda(x)\Psi_{j,k}(x)
dx\right|^2\ .
\eeq
Multiplying Eqs. (13b)--(15b) by
$\psi^+_\lambda(x)\psi_\lambda(x')$ and integrating over $x$ and
$x'$ ($x$ denotes the totality of space and spin variables) we
get
$$
\sum_js^{(\lambda)}_j+\sum_ks^{(\lambda)}_k\ =\ 1 \eqno{(13c)}
$$
$$
\sum_jE_js^{(\lambda)}_j+\sum_kE_ks^{(\lambda)}_k\ =\
\varepsilon_\lambda \eqno{(14c)}
$$
$$
\sum_jE^2_js^{(\lambda)}_j+\sum_kE^2_ks^{(\lambda)}_k\ =\
\varepsilon^2_\lambda+\sigma^2_\lambda \eqno{(15c)}
$$
\beq
\sigma^2_\lambda\ =\int\psi^+_\lambda(x)\Pi(x,x')
\psi_\lambda(x')dxdx'\ .
\eeq
It is remarkable that in contrast to the widths of the
Landau--Migdal quasiparticles \cite{5} the dispersion
$\sigma_\lambda$, Eq.(19), depends upon the wave function
$\psi_\lambda(x)$ rather than the energy $\varepsilon_\lambda$,
thus being roughly the same for all doorway states. In such
situation it is reasonable to identify $\sigma$ with the largest
observed width value. The latter is the widths of the peaks in
the cross sections of quasi-elastic knockout reactions $(p,2p)$
and $(p,pn)$ \cite{7,8} leading to the $1s_{1/2}$ hole states.
According to the above references it is about 20 MeV in all
nuclei.

As seen from Eq.(14c) the doorway state energies $\varepsilon
_\lambda$ are expressed through the energies and $s$-factors of
the actual nuclear states. In general case the latter ones
belong to both the $A-1$ and $A+1$ nuclei, and therefore the
$s$-factors from two different reactions, pick up and stripping,
are required. The absolute values of the $s$-factors are,
however, measured with a rather low accuracy because of both the
experimental and theoretical ambiguities. For this reason the
energies of weakly bound states with $|\varepsilon_\lambda|
<\sigma$ are yet unknown (one should bear in mind that the
low-\-lying states of $A\mp1$ nuclei are Landau--Migdal
quasiparticles \cite{5} rather than the states of nucleon in
static nuclear field).

The situation is more favourable for the states with
$|\varepsilon_\lambda|>\sigma$. In this case, see Eqs. (3) and
(4), the actual nuclear states, over which the doorway ones are
distributed, belong mainly to either the $A-1$ nucleus or the
$A+1$ one, only one term in the lhs (the first for hole states
and the second for particle ones) of Eqs.(13c)--(15c) thus
being active. This is just the case for the strongly bound hole
states which are excited in the quasielastic knockout reactions
$(p,2p)$ and $(p,pn)$~~ \cite{7,8} For this reason the average
energies of the peaks in the cross sections may be identified
with the doorway state energies within the experimental accuracy
of 2--3 MeV. We use the facts that the cross section of the
quasielastic knockout reaction leading to the fixed nuclear
state is proportional to the $s$-factor of this state, and the
absolute values of the $s$-factors are unnecessary when all
states, over which the doorway one is distributed, belong to the
same nucleus (in this case the relative values are sufficient).

The experimental data of Refs.\cite{7,8}, which are used in the
present work, are not free of the following possible ambiguity:
the energy of the knocked-out nucleon is only about 100 MeV in
the experiments. This may be insufficient to neglect the
final--state inelastic interactions leading to an additional
excitation of the final nucleus. As a result of such excitations
the average energies of the peaks may be shifted from the
doorway ones because the reaction mechanism is not a pure
quasielastic knockout in this case. For a greater confidence the
additional quasielastic knockout experiments $(p,p'N)$ or
$(e,e'N)$ are desired, in which the energy of the knock-\-out
nucleon would be of order of 0.5--1 GeV. We hope that our work
will stimulate such experiments.

\section{The static field of nucleus}

Consider the high-energy asymptotics of the Feynman diagrams
constituting the mass operator. Let us begin with those of first
order with respect to the free--space $NN$ interaction. The
Hartree diagrams of Fig.1 are obviously energy-\-independent.
But this is not the case for the corresponding Fock diagrams
resulting from the two-\-particle, Fig.2a, three-\-particle,
Fig.2b, four-\-particle forces, Fig.2c, etc.
\begin{figure}
\centerline{\vspace{0.2cm}\hspace{0.1cm}\epsfig{file=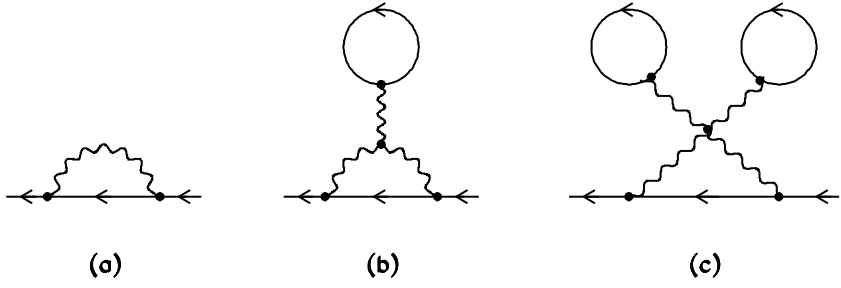,width=12cm}}
\vspace*{-0.5cm}
\caption{First-order exchange contributions to the mass operator.}
\end{figure}
Indeed, according
to the contemporary ideas the $NN$ interaction proceeds via the
exchange by either mesons in the Yukawa-\-like models (OBE
\cite{9}, {Paris \cite{10}, Bonn \cite{11}, OSBEP \cite{12}) or
quarks and gluons in more sophisticated ones. In any case the
interaction includes both the momentum and the energy transfer.
As a result of the latter the Fock diagrams have the
$\varepsilon^{-1}$ asymptotics. Let us demonstrate this for the
diagram of Fig.2a,
\beq
M_F(x,x';\varepsilon)\ =\ \int\frac{id\omega d^3{\bf
q}}{(2\pi)^4}\ e^{i{\bf q}({\bf r}-{\bf r}')} v(q,\omega)
G(x,x'; \varepsilon+\omega)\ ,
\eeq
using the Bonn $B$ potential \cite{11} for the two-particle $NN$
forces. It is the sum of the terms
\beq
v_i(q,\omega)\ =\ g^2_i\left(\frac{\Lambda^2_i-\mu^2_i}{
\Lambda^2_i+q^2-\omega^2}\right)^{2\alpha}
\frac1{\mu^2_i+q^2-\omega^2}\ , \quad
i=\pi,\eta,\rho,\omega,\sigma1,\sigma0,\delta
\eeq
in the four-momentum space, the form of the meson-nucleon
vertices and the sign being specified by the Lorentz symmetry of
the mesons. Both the sign and the Lorentz structure are
disregarded here because they are irrelevant for the energy
dependence. Confining ourselves by the monopole formfactor,
$\alpha=1$, we get
\begin{eqnarray}
&& M_F(x,x';\varepsilon)\ =\ g^2\int\frac{d^3{\bf q}}{(2\pi)^3}
e^{i{\bf q}({\bf r}-{\bf r}')}\left\{\frac1{2\omega_\mu(q)}\quad
\times \right.  \nonumber\\
\times &&\left[\sum_j\frac{\Psi_j(x)\Psi^+_j(x')}{\varepsilon-E_j
+\omega_\mu(q)}+\sum_k\frac{\Psi_k(x)\Psi^+_k(x')}{\varepsilon
-E_k-\omega_\mu(q)}\right] -
\left(1-(\Lambda^2-\mu^2)\frac\partial{\partial\Lambda^2}\right)
\nonumber\\
\times && \left.
\frac1{2\omega_\Lambda(q)}\left[\sum_j
\frac{\Psi_j(x)\Psi^+_j(x')}{\varepsilon-E_j+\omega_\Lambda(q)}
+\sum_k\frac{\Psi_k(x)\Psi^+_k(x')}{\varepsilon-E_k-
\omega_\Lambda(q)}\right]\right\}\\
&& \omega_\mu(q)\ =\ \sqrt{\mu^2+q^2}\ , \quad
\omega_\Lambda(q)\ =\ \sqrt{\Lambda^2+q^2}\ . \nonumber
\end{eqnarray}
In the $\varepsilon\to\infty$ limit this gives
\beq
M_F(x,x';\varepsilon)\ =\ \frac{g^2}{4\pi^2}
\frac{\delta(x-x')}\varepsilon\int
q^2\left[\omega^{-1}_\mu(q)-\omega^{-1}_\Lambda(q)-
\frac{\Lambda^2-\mu^2}2\omega^{-3}_\Lambda(q)\right]dq\ .
\eeq

\begin{figure}
\centerline{\vspace{0.2cm}\hspace{0.1cm}\epsfig{file=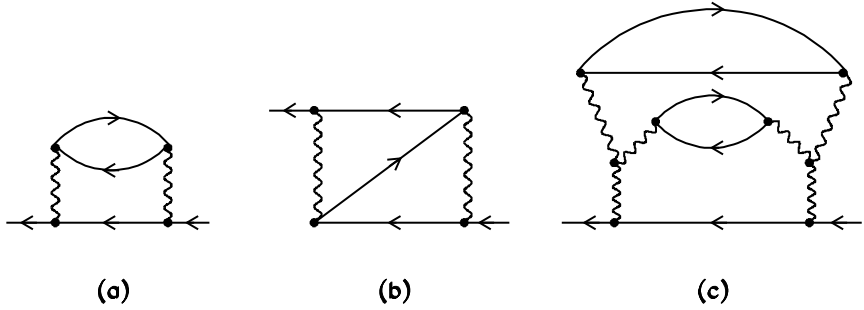,width=12cm}}
\vspace*{-0.5cm}
\caption{ Some second-order diagrams for the mass operator.}
\end{figure}

The second-order diagrams of Fig.3 as well as the higher-\-order
ones contain the propagators of intermediate states, and
therefore they all have at least the $\varepsilon^{-1}$
asymptotics. So the only contribution to the nuclear static
field, Eq.(9), is provided by the Hartree diagrams.

The two-particle contribution to the static field of nucleus,
Fig.1a, is calculated with two different models for the
free-\-space two-\-particle $NN$ interaction, both being of
clear physical meaning and containing small number of adjustable
parameters. The first, the Bonn \cite{11}, is the sum of the OBE
potentials with the vertex formfactors, Eq.(21). The parameters
are adjusted to reproduce the results of the full form of the
Bonn potential which has only one adjustable parameter: see
Ref.\cite{13} for details. In the second, the OSBEP \cite{12},
mesons are treated as objects of nonlinear theory. The mesons
are the same as those in the Bonn $B$, but the form of the
momentum space potentials is different. It is (we have taken
into account that there is no energy transfer in the Hartree
diagrams, i.e. $\omega=0$)
\begin{eqnarray}
 v_i(q)& =& g^2_i \sum^\infty_{n=0}
 \frac{ (2pn+1)^{2pn-2}(1+2(1-p)n)^2(S+1)^n
\alpha^{2n}_\pi\mu^{2pn}_\pi }{ \left\{
1+(4(p+1)\alpha_\pi)^{
-\frac2p}(S+1)^{-\frac1p}\frac{q^2+(2pn+1)^2\mu^2_i}{
4\mu^2_\pi(2pn+1)^2}\right\}^{(S+1)(2pn+1)}} \nonumber\\
&& \times\ \quad \frac1{q^2+(2pn+1)^2\mu^2_i}\ ,
\end{eqnarray}
where $p=1/2$ for scalar mesons and $p=1$ for pseudoscalar and
vector ones, $S$ is the spin of the meson, and the sum over $n$
is practically converging at $n=4$~ \cite{12}.

Both these approaches permit one to check the status of the
Walecka model \cite{14} by calculating the  values of the vector
and scalar fields in nuclear matter. For the case of
charge-\-symmetric matter
\beq
V\ =\ v_\omega(0)\rho\ , \quad S\ =\ -\left(\frac34\
v_{\sigma1}(0)+\frac14\ v_{\sigma0}(0)\right)\rho_s\ ,
\eeq
where the scalar density $\rho_s$ is
\beq
\rho_s\ =\ \rho-\frac{2\tau}{(2m+S-V)^2}\ , \quad \tau\ =\
\frac35\ k^2_F\rho\ ,
\eeq
$k_F$ is the Fermi momentum and $m$ is the free nucleon mass.
Using the conventional equilibrium value of the nuclear matter
density, $\rho=0.17\ $fm$^{-3}$, and the parameters of Table 5
of Ref.\cite{11} and Table 1 of Ref.\cite{12} we get
\beq
V=+284 \mbox{ MeV }, \quad S=-367\mbox{ MeV}
\eeq
for the Bonn $B$ potential and
\beq
V=+322 \mbox{ MeV }, \quad S=-404 \mbox{ MeV }
\eeq
for the OSBEP, both being close to those provided by the Dirac
phenomenology \cite{15}. So the contemporary $NN$ interaction
potentials lead to nuclear relativity, the latter thus being
really existing phenomenon rather than the suggestion of
J.D.~Walecka.

For this reason the doorway state wave functions
$\psi_\lambda(x)$ should be treated as Dirac bispinors obeying
the Dirac equation with
\beq
{\cal H}_{sp}\ =\ -i\gamma_0\mbox{\boldmath$\gamma\nabla$}
+i\Phi(r)\mbox{\boldmath$\gamma$}\ \frac{\bf r}r+ (\gamma_0-1)
m+V(r)+\gamma_0S(r)\ .
\eeq
The scalar and vector fields of finite nuclei consist of the
isoscalar and isovector parts, the vector field also including
the Coulomb potential
\begin{eqnarray}
&& \hspace{-0.8cm}
S(r)=S_0(r)-\tau_3S_1(r),\ \tau_3=\left\{{-1,\ n \atop +1,\
p}\right|,\ V(r)=V_\omega(r)-\tau_3V_\rho(r)+\frac{1+\tau_3}2
C(r), \nonumber\\
&& S_0(r)\ =\ -\int\left(\frac34\ v_{\sigma1}(q)+\frac14\
v_{\sigma0}(q)\right)F_s(q)\ e^{i{\bf qr}}\frac{d^3\bf
q}{(2\pi)^3}\ , \nonumber\\
&& S_1(r)\ =\ -\int\left(v_\delta(q)+\frac14v_{\sigma1}(q)
-\frac14v_{\sigma0}(q)\right)F^-_s(q)e^{i\bf qr} \frac{d^3\bf
r}{(2\pi)^3} \\
&& V_\omega(r)\ \int v_\omega(q)F(q)e^{i\bf qr} \frac{d^3\bf
q}{(2\pi)^3}\ , \quad V_\rho(r)\ =\int v_\rho(q)F^-(q) e^{i\bf
qr} \frac{d^3\bf q}{(2\pi)^3}\ , \nonumber\\
&& C(r)\ =\ e^2\int\frac{\rho_{ch}(r')}{|{\bf r}-{\bf r}'|}\ d^3
{\bf r}'\ ;\ \quad \Phi(r)\ =\ \tau_3\frac\kappa{2m}\
\frac{dV_\rho}{dr}\ , \nonumber
\end{eqnarray}
where
\begin{eqnarray}
&& F_s(q)=\int\rho_s(r)e^{-\bf qr} d^3{\bf r},\quad F^-_s(q)
=\int\rho^-_s(r)e^{-i\bf qr}d^3{\bf r} \nonumber\\
&&  \hspace{-1cm}
F(q)=\int\rho(r)e^{-i\bf qr}d^3{\bf r}, \; F^-(q)=\int
\left(\rho^-(r)+\frac\kappa{2mr^2}\frac d{dr}(r^2w^-(r)) \right)
e^{-i\bf qr}d^3{\bf r} \nonumber\\
&& \rho(r)=\rho_n(r)+\rho_p(r),\; \rho^-(r)=\rho_n(r)-\rho_p(r),
\; \rho_s(r)=\rho_{sn}(r)+\rho_{sp}(r)\ , \nonumber \\
&& \rho^-_s(r)=\rho_{sn}(r)-\rho_{sp}(r)\ , \quad
w^-(r)=w_n(r)-w_p(r)\ .
\end{eqnarray}
The scalar densities and the quantities $w(r)$ are
\begin{eqnarray}
\rho_s(r) &=& \rho(r)-\frac{2(\tau(r)+\Phi(r)\rho'(r)
+\Phi^2(r)\rho(r))}{(2m+S(r)-V(r))^2}\\
w(r) &=& \frac{\rho'(r)+2\Phi(r)\rho(r)}{2m+S(r)-V(r)}\ .
\end{eqnarray}
They are calculated separately for neutrons and protons. The
quantity $\tau(r)$ is calculated in the local density
approximation using Eq.(26). The isovector quantity $\Phi(r)$,
Eqs.(29) and (30), arises from the tensor $\rho NN$ coupling,
$\kappa=f_\rho/g_\rho$ is the tensor-\-to-\-vector  coupling
ratio.

So the two-particle contributions may be determined from
experiment by using a definite model for the two-\-particle $NN$
interaction. Little is known, however, about the many-\-particle
$NN$ forces. In such conditions it is reasonable to look for the
many-\-particle contribution as a power series expansion over the
nucleon density distribution:
\beq
U_m(r)\ =\ S_m(r)+V_m(r)\ =\ a_3\rho^2(r)+a_4\rho^3(r)+\ \cdots
\eeq
the $\rho^2(\rho^3)$ term resulting from three-
(four)-\-particle forces etc. To elucidate the physical meaning
of the coefficients let us consider a general form of the
three-\-particle term:
\beq
U_3(r)\ =\ \int f_3\left(|{\bf r}-{\bf r}_1|, |{\bf r}-{\bf
r}_2| \right) \rho(r_1)\rho(r_2)d^3{\bf r}_1d^3{\bf r}_2\ .
\eeq
In the homogeneous nuclear matter this gives
\beq
 U_3\ =\ \rho^2\int f_3\left(|{\bf r}\cdot{\bf r}_1|, |{\bf
r}-{\bf r}_2| \right)d^3{\bf r}_1d^3{\bf r}_2\ , \eeq and
therefore \beq a_3\ =\ \int
f_3(\eta,\xi)d^3\mbox{\boldmath$\xi$} d^3
\mbox{\boldmath$\eta$}\ .
\eeq
In the same way
\beq
a_4\ =\ \int f_4(\xi,\eta,\zeta) d^3\mbox{\boldmath$\xi$}
d^3\mbox{\boldmath$\eta$}d^3\mbox{\boldmath$\zeta$}\ .
\eeq
These volume integrals are the only parameters which do not
require any specific model for the many-\-particle $NN$ forces.
Such  model is, however, necessary to take into account the
finite range of the forces. We did not try to do this since
(a)~the problem of many-\-particle $NN$ interaction mechanism is
beyond the scope of our work and (b)~the additional adjustable
parameters describing the finite range cannot be safely
determined because of the insufficient accuracy of the available
experimental data.

The same reason forced us to introduce as little free parameters
as possible and use all permissible simplifications. In
particular, the many-\-particle terms are assumed to be equally
distributed between the scalar and vector fields:
\beq
S_m(r)\ =\ V_m(r)\ =\ \frac12\ U_m(r)\ .
\eeq

\section{Results}

The observed and calculated spectra of the doorway state
energies in $^{40}$Ca, $^{90}$Zr and $^{208}$Pb nuclei are
plotted in Figs. 4 and 5.

\begin{figure}
\centerline{\vspace{0.2cm}\hspace{0.1cm}\epsfig{file=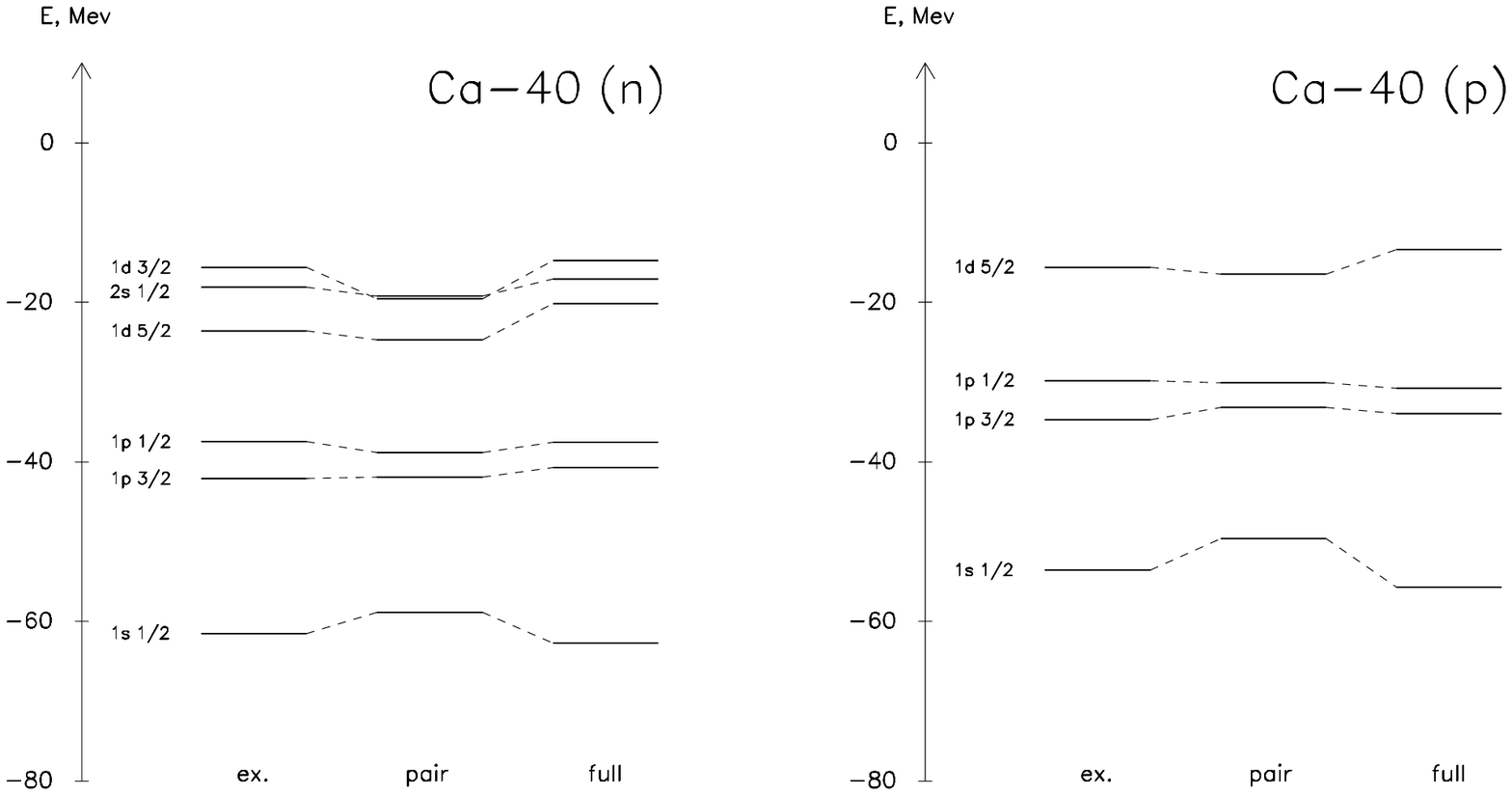,width=12cm}}
\vspace*{-0.5cm}
\centerline{\vspace{0.2cm}\hspace{0.1cm}\epsfig{file=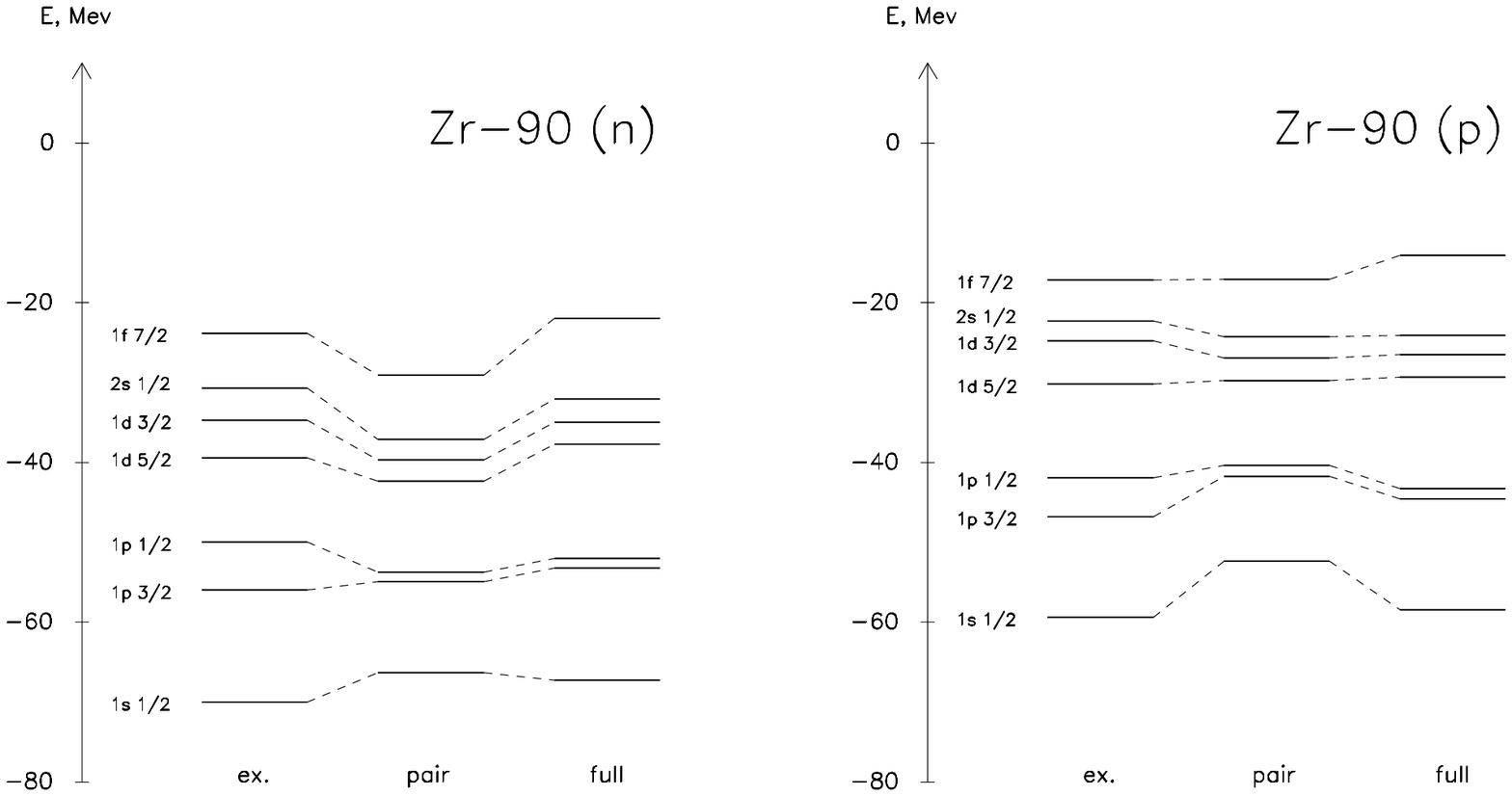,width=12cm}}
\vspace*{-0.5cm}
\centerline{\vspace{0.2cm}\hspace{0.1cm}\epsfig{file=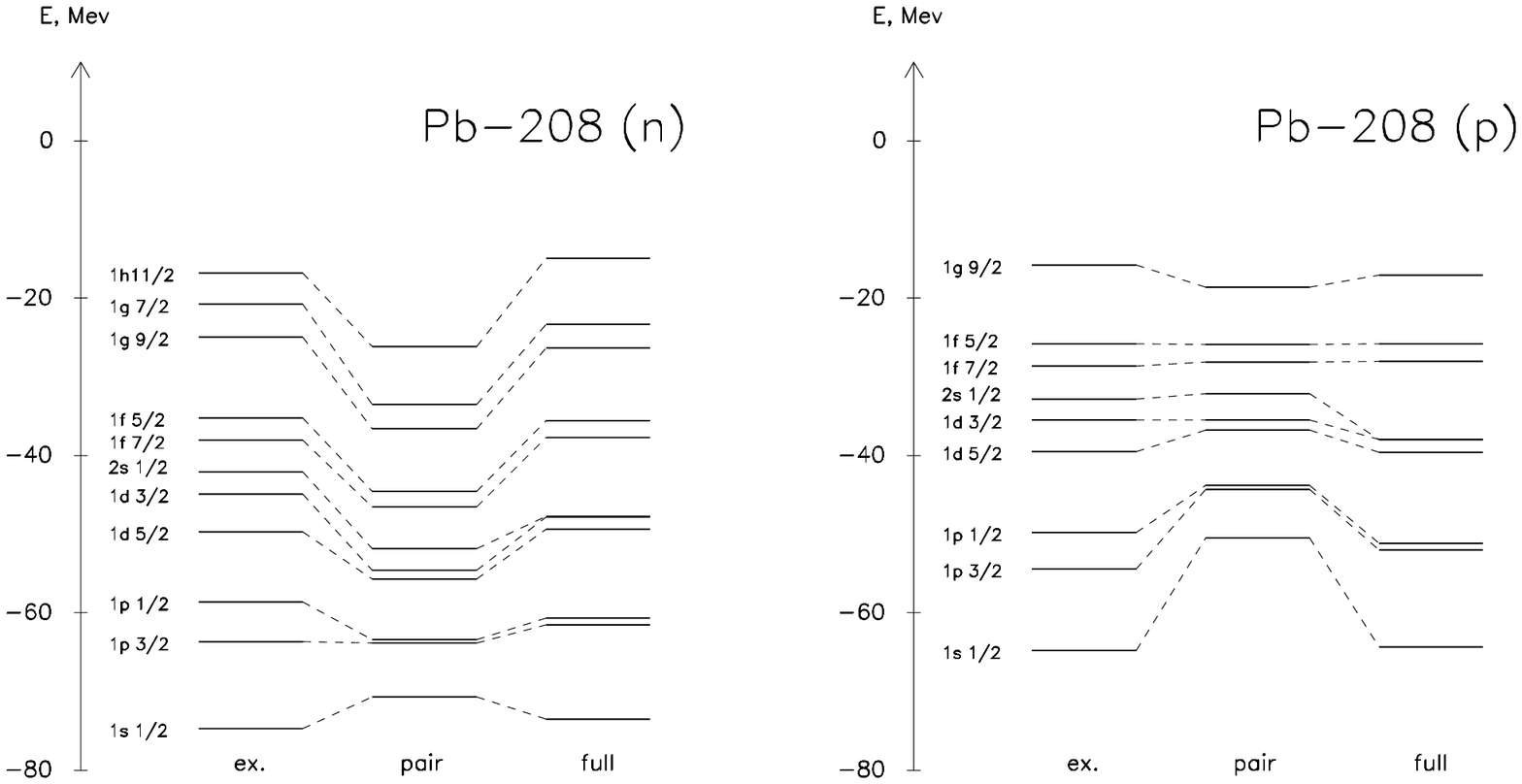,width=12cm}}
\vspace*{-0.5cm}
\caption{Spectra of doorway state energies with the Bonn B
potential for the two-\-particle forces.          }
\end{figure}
The calculations are performed with
two different two-\-particle potentials: the Bonn $B$, Fig.4,
and the OSBEP, Fig.5.

\begin{figure}
\centerline{\vspace{0.2cm}\hspace{0.1cm}\epsfig{file=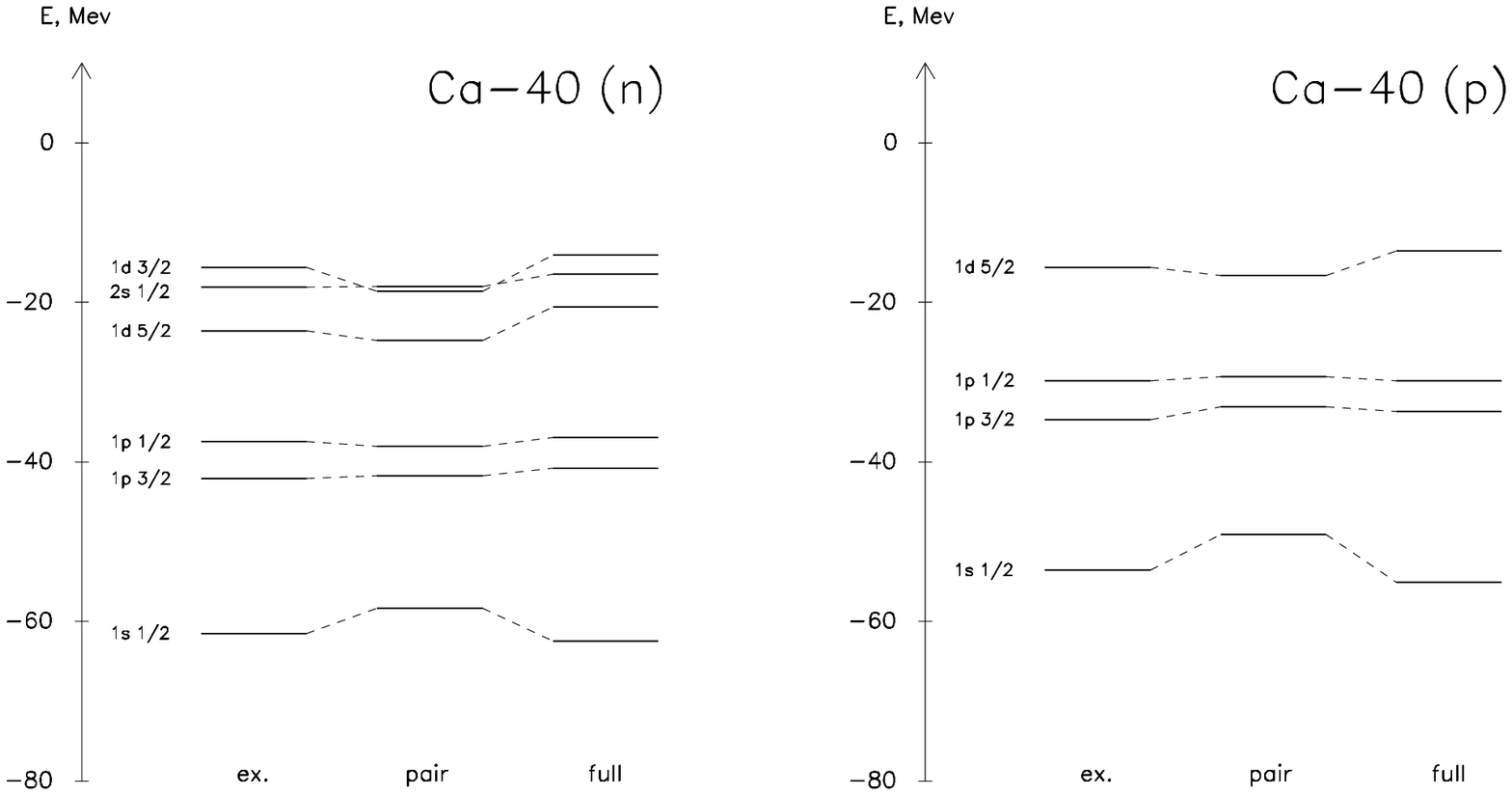,width=12cm}}
\vspace*{-0.5cm}
\centerline{\vspace{0.2cm}\hspace{0.1cm}\epsfig{file=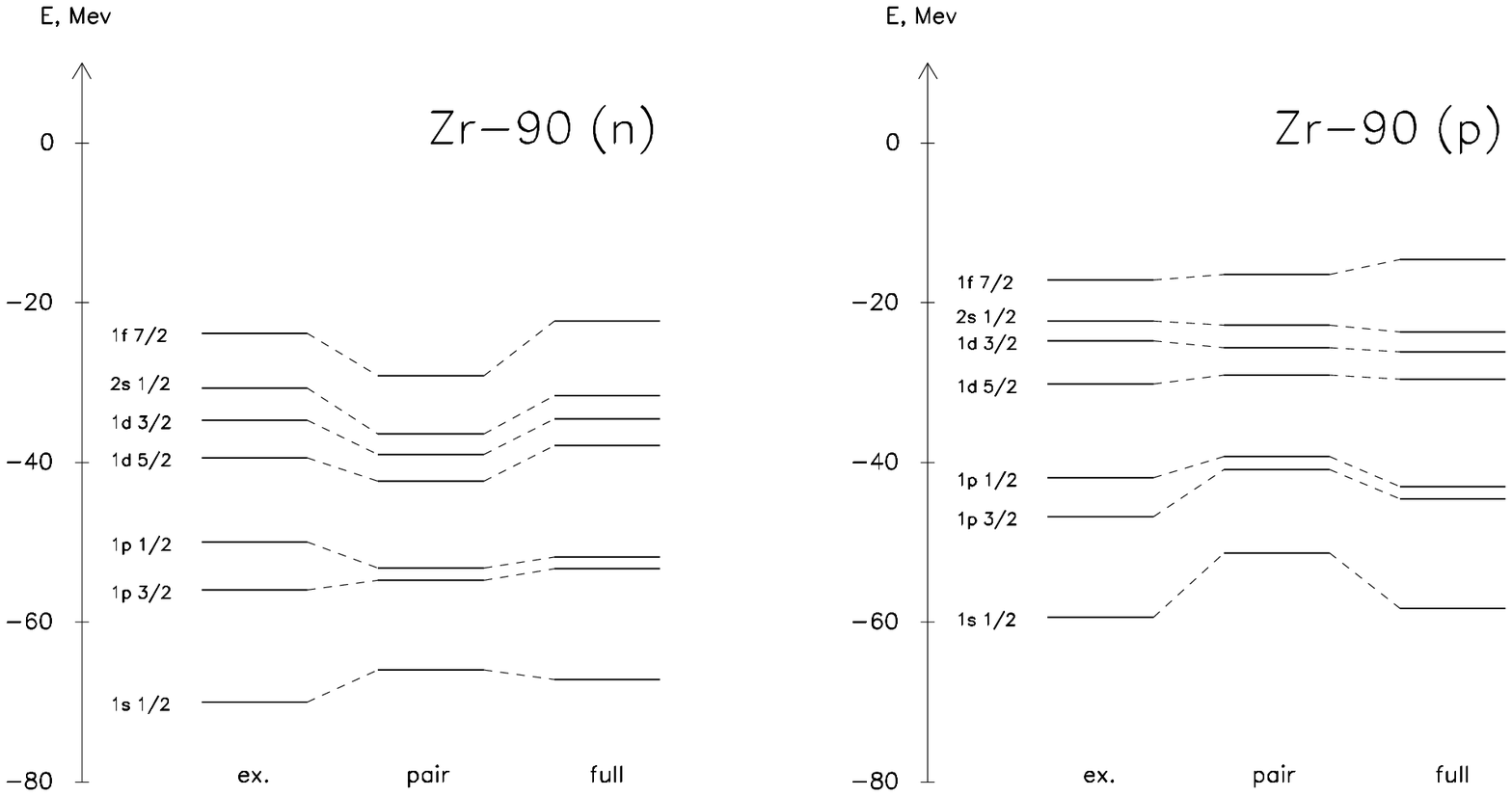,width=12cm}}
\vspace*{-0.5cm}
\centerline{\vspace{0.2cm}\hspace{0.1cm}\epsfig{file=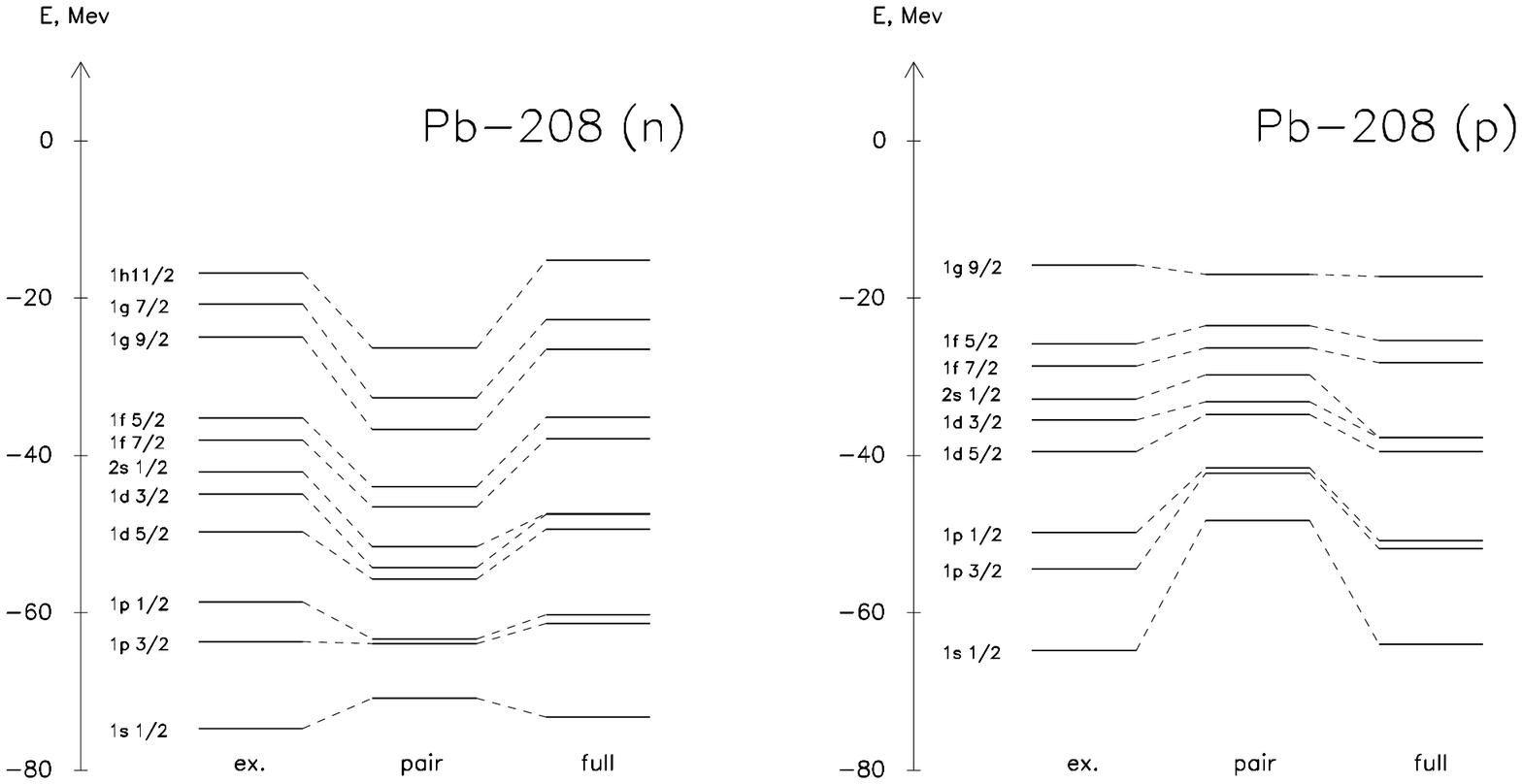,width=12cm}}
\vspace*{-0.5cm}
\caption{ The same with the OSBEP. }
\end{figure}
The results for the two-\-particle forces
only are labelled as "pair". As seen from the figures the "pair"
spectra are compressed compared to the observed ones, the lowest
$1s_{1/2}$ states being significantly underbound. This means
that the potential well resulting from the two-\-particle forces
only is too wide but insufficient deep, and so the actual well
must be deeper and narrower as illustrated by Fig.6.

\begin{figure}
\centerline{\vspace{0.2cm}\hspace{0.1cm}\epsfig{file=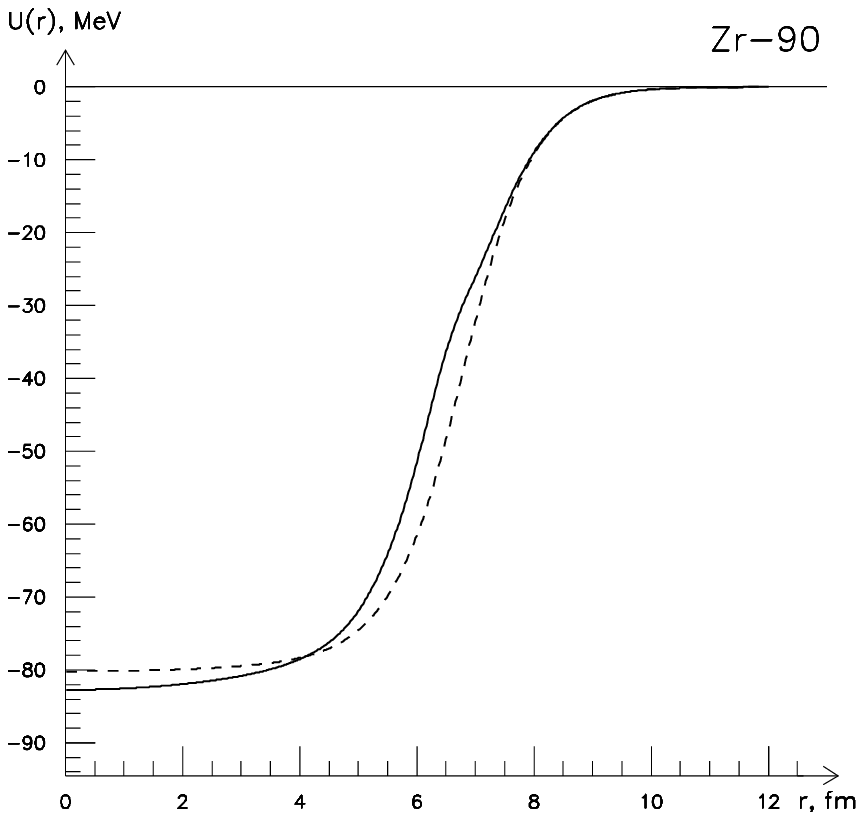,width=12cm}}
\vspace*{-0.5cm}
\caption{Isoscalar part of the static field in $^{90}$Zr.
The dashed  and full curves are for the "pair" and actual wells
respectively. The calculations are performed with the Bonn B
two-\-particle forces and original nucleon density distributions
of Ref. [3].}
\end{figure}
Hence, the
many-\-particle contribution (as discussed above this is the
only reason for the difference between the actual and "pair"
wells) consists of the attractive and repulsive parts, the
radius of the former being less than that of the latter. The
most simple form obeying this condition is provided by the sum
of first two terms of the expansion (34) with $a_3>0$ and
$a_4<0$. In other words, the free-\-space many-\-particle $NN$
interaction includes at least the three-\-particle repulsion and
the four-\-particle attraction (of course the presence of higher
many-\-particle forces is not excluded).

Accounting for the fact that the many-particle forces contribute
to both the isoscalar and isovector parts of the static nuclear
field the quantity $U_m(r)$ is chosen in the form
\begin{eqnarray}
U_m(r) &=& a_3\rho^2(r)+a_4\rho^3(r)-\tau_3\left[a^-_3\rho(r)
+a^-_4\rho^2(r)\right]\rho^-(r) \nonumber\\
\rho(r) &=& \rho_n(r)+\rho_p(r)\ , \quad \rho^-(r)\ =\
\rho_n(r)-\rho_p(r)\ .
\end{eqnarray}
The finite size of nucleon is taken into account in the
free-space $NN$ forces, and therefore the static field of
nucleus is expressed through the point nucleon densities. The
proton ones $\rho_p(r)$ are obtained from the charge density
distributions of Ref.\cite{2} by a usual deconvolution
procedure. They are shown in Fig. 7a. The point neutron
densities $\rho_n(r)$ are obtained from the folded densities of
Ref.\cite{3} in the same way.

\begin{figure}
\centerline{\vspace{0.2cm}\hspace{0.1cm}\epsfig{file=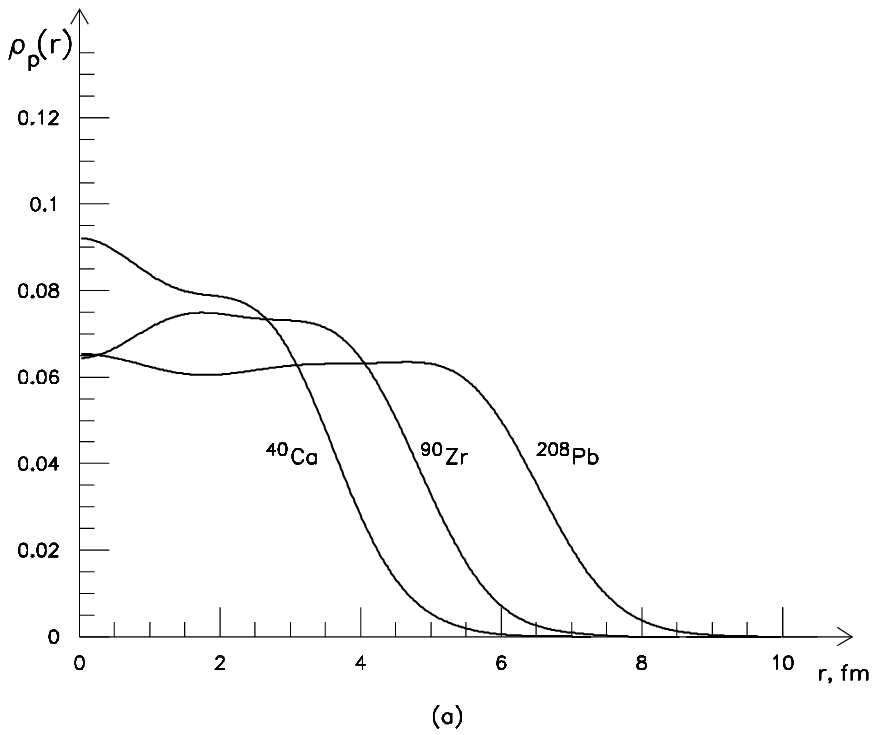,width=9cm}}
\vspace*{-0.5cm}
\centerline{\vspace{0.2cm}\hspace{0.1cm}\epsfig{file=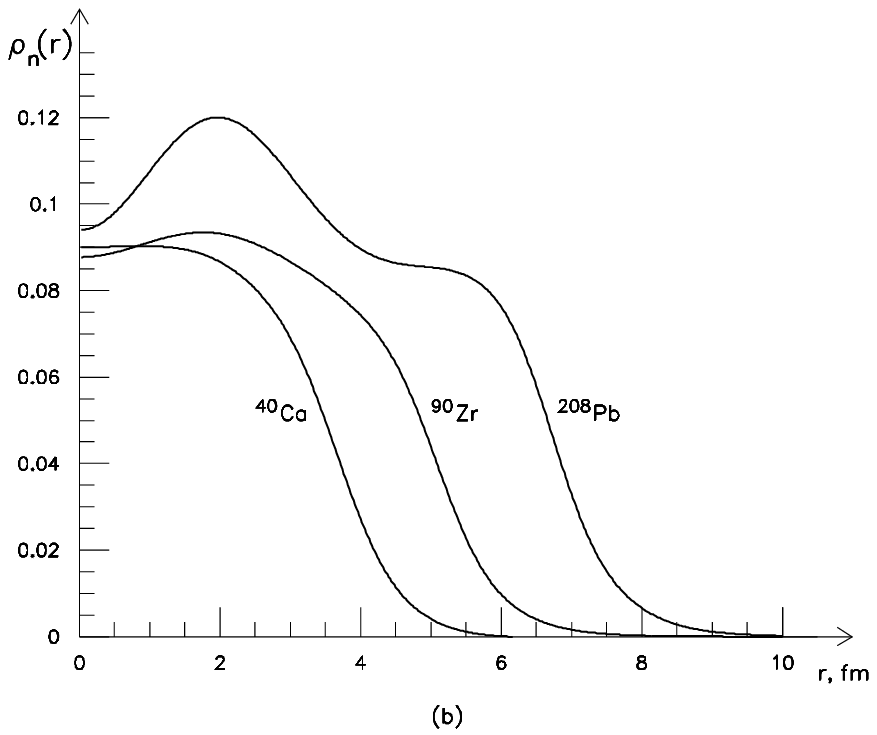,width=9cm}}
\vspace*{-0.5cm}
\caption{ Density distributions of protons (a) and neutrons
(b) in $^{40}$Ca, $^{90}$Zr and $^{208}$Pb nuclei.}
\end{figure}

The data of Ref. \cite{2} are based on high precision
measurements of elastic electron-\-nucleus scattering thus
providing the proton density distributions in the whole nuclear
region. The situation for the neutron densities is different
since the elastic 1 GeV proton-\-nucleus scattering underlying
the data of Ref. \cite{3} is sensitive mainly to the surface
region of nucleus because of the absorption. For this reason the
neutron density distributions $\rho_n(r)$ may differ from the
Woods--Saxon--like ones of Ref. \cite{3} in nuclear interior (as
seen from Fig. 7a the proton densities are indeed different from
the Woods--Saxon-\-like ones). The latter is just the region to
which the doorway state energies are sensitive, and therefore
they may be used to specify the Ref. \cite{3} data for the
neutron densities. We looked for the latter ones in the form
\beq
\rho_n(r)\ =\ \rho_0\bigg[W_A(r)+\alpha W_A(0)\varphi_4(\beta
r)\bigg]\ ,
\eeq
where $W_A(r)$ are the deconvoluted neutron densities of Ref.
\cite{3} and $\varphi_4(x)$ is the fourth Hermite function. The
neutron density parameters $\alpha,\beta$ and the strength ones
$a_3,a_4,a^-_3,a^-_4$ are determined from the best fit for both
the doorway state energies and the elastic 1 GeV
proton-\-nucleus scattering, the latter being calculated within
the Glauber theory \cite{16}.

 \begin{table}
\caption{Neutron density parameters}
	 \begin{center}
\bigskip
\begin{tabular}{||c|c|c||c|c||} \hline\hline
& \multicolumn{2}{c||}{Bonn $B$} & \multicolumn{2}{c||}{OSBEP}\\
\cline{2-5}
& $\alpha$ & $\beta$ & $\alpha$ & $\beta$\\ \cline{2-5}
$^{40}$Ca & $-0.0295$ & 0.5314 & $-0.0255$ & 0.5230\\
$^{90}$Zr & $-0.0758$ & 0.5551 & $-0.0646$ & 0.5442\\
$^{208}$Pb & $-0.2645$ & 0.5445 & $-0.2667$ & 0.5389\\
\hline\hline
\end{tabular} \end{center}\end{table}

The density parameters are shown in Table 1. They are different
for the two choices of the two-\-particle forces, but the
difference is rather small. For this reason neither the
resulting neutron density distributions, Fig. 7b, nor the 1 GeV
proton--nucleus elastic scattering cross sections, Fig.8, are
distinguishable in the figures. We also calculated the
proton--nucleus cross sections with the original results of
Ref.\cite{3} for the density distributions.
\begin{figure}
\centerline{\vspace{0.2cm}\hspace{0.1cm}\epsfig{file=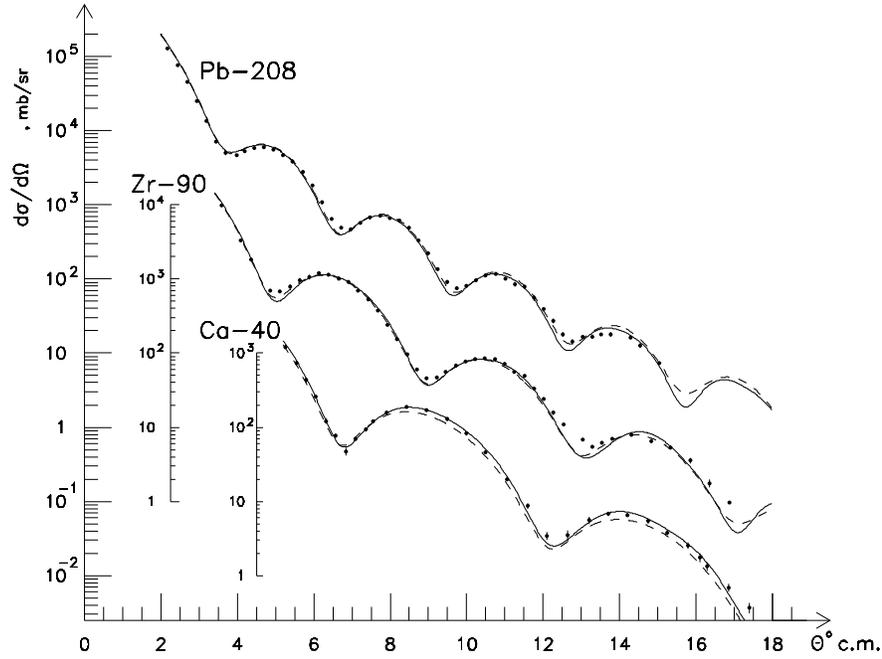,width=12cm}}
\vspace*{-0.5cm}
\caption{1 GeV proton-nucleus elastic scattering cross
sections. The dashed  and full curves are calculated with the
original neutron density distributions of Ref. [3] and the
specified ones respectively.}
\end{figure}
As seen from Fig.8
the agreement with experiment is equally good for both the
specified densities, Eq.(41), and the original ones. The
many--particle strength parameters are
\begin{eqnarray}
a_3=16.9296 \mbox{ fm}^5\ , && a_4=-107.6744 \mbox{ fm}^8\ ,
\nonumber\\
a^-_3=25.5873\mbox{ fm}^5\ , && a^-_4=-128.5134\mbox{ fm}^8
\end{eqnarray}
for the Bonn $B$ two-particle forces and
\begin{eqnarray}
a_3=17.0011 \mbox{ fm}^5\ , && a_4=-110.3747 \mbox{ fm}^8\ ,
\nonumber\\
a^-_3=26.9036\mbox{ fm}^5\ , && a^-_4=-130.1210\mbox{ fm}^8
\end{eqnarray}
for the OSBEP ones. As seen from Eqs. (42) and (43) the strength
parameters of the free-\-space many-\-particle forces are almost
the same for the two cases. This is not surprising because both
the Bonn $B$ and the OSBEP potentials provide an equally good
description of the two-\-nucleon data, see the discussion in the
Introduction.

The results including both the contribution from the
many-\-particle forces and the specified neutron densities are
labelled as "full" in Figs. 4 and 5. The "full" doorway state
energies agree with the observed ones (which are labelled as
"exp") within the experimental error of 3 MeV. The exception is
provided by the $2s_{1/2}$ states in $^{208}$Pb: in this case
the discrepancy is about 5 MeV. The reason is not clear yet, but
the discrepancy does not exceed two experimental errors.

To estimate the relative importance of the two-particle and
many-particle contributions to the static field of nucleus let
us perform the calculations for nuclear matter, see Sect.3.
First consider the isoscalar part. The two-\-particle
contribution is
\beq
U_2=V_2+S_2=\hbar c\left[v_\omega(0)\rho-\left(\frac34
v_{\sigma1}(0)+\frac14v_{\sigma0}(0)\right)\rho_s\right]=
\left\{ {-83\mbox{ MeV},\ \mbox{ Bonn }B \atop -82\mbox{ MeV},\
\mbox{ OSBEP },} \right.
\eeq
whereas those from three-particle and four-particle forces are
\begin{eqnarray}
U_3 &=& \hbar ca_3\rho^2=\left\{ {96.5 \mbox{ MeV,~~~ Bonn }B
\atop 97\mbox{ MeV,~~~ OSBEP} } \right. \nonumber\\ U_4 &=&
 \hbar ca_4\rho^3=\left\{ {-104 \mbox{ MeV,~~  Bonn }B \atop
-107\mbox{ MeV,~~  OSBEP}} \right.
 \end{eqnarray}
the
many-particle contributions thus being as large as the
two-particle one.

The isovector part may be estimated by putting
$\rho^-=\frac{N-Z}A\rho$ and $\rho^-_s=\frac{N-Z}A\rho_s$. The
two-particle contribution is (see Sect.3)
\begin{eqnarray}
U^-_2 &=&\hbar c\left[v_\rho(0)\rho^--\left(v_\delta(0)+\frac14
v_{\sigma1}(0)-\frac14v_{\sigma0}(0)\right)\rho^-_s\right]\ =
\nonumber\\
&=&
{6\frac{N-Z}A \mbox{ MeV,~~~  Bonn }B \atop 0.15\frac{N-Z}A
\mbox{ MeV,~  OSBEP} }
\end{eqnarray}
the many-particle one being
\beq
U^-_m=\hbar c\left(a^-_3\rho+a^-_4\rho^2\right)= \left\{
\begin{array}{ll}
\displaystyle (146-125=21)\frac{N-Z}A\mbox{ MeV,} &\mbox{ Bonn
}B\\ \displaystyle (153-126=27)\frac{N-Z}A\mbox{ MeV,} &
\mbox{ OSBEP }.
\end{array} \right.
\eeq
So the many-particle
forces provide the dominant part of the isovector nuclear
potential.  The reason is due to the fact that the
two-\-particle contribution arises from the exchange by
isovector mesons $\rho$ and $\delta$ which are weakly coupled to
nucleon, see Table 5 of Ref.\cite{11} and Table 1 of
Ref.\cite{12}.

\section{Summary}

The above results give rise to the following general
conclusions:

1. Our results for the many-particle forces are quite
competitive with those from the few-nucleon systems \cite{17}.
Indeed, the properties of the latter ones (binding energies,
sizes, formfactors etc.) are expressed through the interaction
in all orders of the perturbation theory, and therefore the
solution of a rather complicated quantum mechanical problem is
necessary to get the information on the many-\-particle forces.
In contrast to the few-nucleon systems the doorway states for
the one-\-nucleon transfer reactions in complex nuclei are
solutions of a much more simple problem for one nucleon in
central field. In addition the static nuclear field is expressed
through the $NN$ forces in first order of the perturbation
theory, the results thus being very visual, see Figs. 1 and 6.
The information from the doorway states is, however, restricted
because it concerns only spin-\-independent terms of the
many-\-particle forces (the spin-\-dependent ones do not
contribute to the Hartree diagrams). Nevertheless it is a useful
addition to that from the few-\-nucleon systems.

Two important points should be mentioned in this connection.
(i)~Only three-\-particle forces (in addition to the
two-\-particle ones) are included in all available calculations
for the few-\-nucleon systems. Our results clearly show that
this is insufficient. (ii)~Calculating the nuclear correlation
effects (binding energies and rms radii of finite nuclei,
equation of state of nuclear matter, etc.), with the free-\-space
$NN$ interaction there is no reason to neglect the
many-\-particle forces because they are as strong as the
two-\-particle ones, compare Eqs. (44) and (45).

2. The effective three-particle and four-particle forces are
also repulsive and attractive respectively in the recent
calculations within the relativistic mean-\-field approximation
\cite{18,19}, see the Appendix. Such forces include implicity
the correlation effects which are not taken into account
explicitly within this framework. For this reason the above
signs of the forces might be treated as the artifact of the
approximation. But our results for the free-space
many-\-particle forces show that this is not the artifact.

\def\thesection{Appendix \Alph{section}}
\def\theequation{\Alph{section}.\arabic{equation}}
\setcounter{equation}{0}

\appendix\section{Appendix}
The potential energy of the $\sigma$ mesons is \cite{18,19}
\beq
U(\sigma)\ =\ \frac{\mu^2}2\ \sigma^2+\frac{\lambda_3}3\
\sigma^3+\frac{\lambda_4}4\ \sigma^4
\eeq
with $\lambda_3<0$ and $\lambda_4<0$, the scalar field
$S=g\sigma$ thus obeying the equation
\beq
(\Delta-\mu^2)S\ =\ g^2_\sigma\rho_s+\frac{\lambda_3}g\ S^2
+\frac{\lambda_4}{g^2}\ S^3\ .
\eeq
Let us use the following iteration procedure
\beq
(\Delta-\mu^2)S_n\ =\ g^2_\sigma\rho_s+\frac{\lambda_3}g
S^2_{n-1}+\frac{\lambda_4}{g^2}S^3_{n-1}
\eeq
with
\beq
(\Delta-\mu^2)S_0\ =\ g^2_\sigma\rho_s
\eeq
for the initial iteration. The result is
\begin{eqnarray}
S(r) &=& -g^2\int y(|\br-\br_1|)\rho^{(r)}_s(r_1)d^3\br_1 \quad
+ \nonumber\\
&+& \int
f_3(|\br-\br_1|,|\br-\br_2|)\rho_s(r_1)\rho_s(r_2)d\br_1d\br_2\
+ \\
&+&\int f_4(|\br-\br_1|,|\br-\br_2|,|\br-\br_3|
)\rho_s(r_1)\rho_s(r_2)
\rho_s(r_3)d^3\br_1 d^3\br_2 d^3\br_3\ +\ \cdots\ , \nonumber
\end{eqnarray}
where
\begin{eqnarray}
&&  y(|\br-\br'|)\ =\ \frac{\exp(-\mu|\br-\br'|)}{4\pi
|\br-\br'|}\\
&&\hspace{-1.cm} f_3(|\br-\br_1|,|\br-\br_2|)\ =\
-\lambda_3g^3\int
y(|\br-\br'|)y(|\br_1-\br'|)y(|\br_2-\br'|)d^3\br'\\
&& f_4(|\br-\br_1|,|\br-\br_2|,|\br-\br_3|)\quad = \nonumber\\
&=&\lambda_4g^4\int
y(|\br-\br'|)y(|\br_1-\br'|)y(|\br_2-\br'|)y(|\br_3-\br'|)d^3r'\
- \\
&-& 2\lambda^2_3g^4\int y(|\br-\br'|)y(|\br_1-\br'|)y
(|\br'-\br''|)y(|\br_2-\br''|)y(|\br_3-\br''|)d^3\br'd^3\br''.
\nonumber
\end{eqnarray}
The dots in the rhs of Eq.(A.5) represent the higher-power terms
in respect of $\rho_s$ resulting from the higher many-particle
forces. As seen from (A.7) the three-\-particle force is
repulsive because of the sign of $\lambda_3$ ($g>0$ in
Refs.\cite{18,19}). The four--particle one, Eq.(A.8), consists
of two terms. The first is of first order with respect to the
$\lambda_4$ term of (A.1). It is attractive because of the sign
of $\lambda_4$. The second is of second order with respect to
the $\lambda_3$ term. It is attractive irrespective of the sign
of $\lambda_3$.

The volume integrals of the forces (A.7) and (A.8), Eqs. (37)
and (38), are
\beq
a_3\ =\ -g^3\lambda_3\mu^{-6}, \quad a_4\ =\
g^4(\lambda_4-2\lambda^2_3\mu^{-2})\mu^{-8}\ .
\eeq
The least values of these quantities correspond to the NL-SH
parameter set of Ref.\cite{18}, Table 2 of this reference. They
are
\beq
a_3\ =\ 21.9\mbox{ fm}^5\ , \quad a_4\ =\ -136.5\mbox{ fm}^8\ ,
\eeq
thus being rather close to the free-space values, Eqs. (42) and
(43).

\end{document}